# Millimeter-wave Gbps Broadband Evolution towards 5G: Fixed Access and Backhaul

Zhouyue Pi, Junil Choi, and Robert Heath Jr.

*Abstract* — As wireless communication evolves towards 5G, both fixed broadband and mobile broadband will play a crucial part in providing the Gbps infrastructure for a connected society. This paper proposes a Millimeter-wave Gbps Broadband (MGB) system as the solution to *two* critical problems in this evolution: last-mile access for fixed broadband and small cell backhaul for mobile broadband. The key idea is to use spectrum that is already available in the millimeter wave bands for fixed wireless access with optimized dynamic beamforming and massive MIMO infrastructure to achieve high capacity with wide area coverage. This paper explains the MGB concept and describes potential array architectures for realizing the system. Simulations demonstrate that with 500 MHz of bandwidth (at 39 GHz band) and 28 dBm transmission power (55 dBm EIRP), it is possible to provide more than 11 Gbps backhaul capacity for 96 small cells within 1-km radius.

## I. Introduction

MOBILE computing is one of the greatest innovations in the history of technology. The rapid adoption of smartphones and the explosive growth of data traffic due to these devices has been phenomenal. As the world anticipates more connected devices (*i.e.*, the Internet of Things, vehicle-to-vehicle (V2V) communications, or wearable devices) and more value added applications & services (*e.g.*, ultra high definition video, 360° video, virtual reality, smart cars, etc.), leading industry experts are calling for the fifth generation (5G) networks to provide 1000x capacity increase over 4G. Three technologies have been identified to achieve this goal: millimeter-wave (mmWave) mobile broadband [1], massive MIMO [2], and small cells [3]. The technologies need to be used in combination; no one technology can provide the expected 1000x capacity increase.

Each technology option comes with its own specific challenges. MmWave carrier frequencies need new spectrum allocations. Both mmWave and massive MIMO require new hardware at the base station and global standardization before commercialization. Small cells can be applied even with today's technology, but site acquisition and backhaul become increasingly challenging especially when the small cell need to support gigabit per second physical layer technologies.

Applying mmWave and massive MIMO to 5G mobile broadband involves new system designs. The challenge faced by small cell backhaul, however, is similar to the last-mile problem of fixed broadband access. As society moves towards the Gbps era, broadband service providers face the demanding task of upgrading their *entire* twisted-pair or coaxial cable based infrastructure to fiber. While it may be manageable to upgrade the backbone network, it may take years – if not decades – to upgrade all the wires to millions of homes or enterprises, with the caveat this approach may not be economically viable at all. A scalable way of deploying a network with large number of small cells and fixed broadband access points is crucial for the success of 5G.

In this paper, we propose a millimeter-wave Gbps broadband (MGB) solution that provides Gbps links to small cells and fixed broadband access points. For simplicity, we refer to both mobile broadband small cells (*e.g.*, an LTE small cell) and fixed broadband access points (*e.g.*, a Wi-Fi access point) as small cells. The MGB system provides area coverage of Gbps connections, allowing small cells to be deployed anywhere within the coverage without being limited by access to wired infrastructure. This will significantly increases the Gbps coverage for mobile devices and connected vehicles with minimal latency. In addition, the fixed nature of the small cells in the MGB system provides favorable channel conditions that allow sophisticated transmission to achieve much higher spectral efficiency than that attainable in a mobile environment.

Spectrum is readily available for MGB systems. For example, in the United States, a total of 2.7 gigahertz of bandwidth is available for MGB in the LMDS band (27.5 – 28.35, 29.1 – 29.25, and 31.0 – 31.3 GHz), and the 39 GHz band (38.6 – 40.0 GHz) as these bands are already licensed for fixed point-to-multipoint services on geographic area basis. This means that, unlike the mmWave mobile technologies that need global standards and 5G spectrum before commercialization, the MGB system proposed in this paper can be developed and deployed today.

In both applications as small cell backhaul and as Gbps broadband access, the MGB system is more attractive than fiber or point-to-point microwave solutions. Its flexibility, scalability, and potential for lower capital expenditure and much lower operating costs, make it much easier to deploy and reconfigure networks. This provides cellular operators and broadband service providers a higher return on investment in

J. Choi and R. Heath are with Dept. of Electrical and Computer Engineering at The University of Texas at Austin. Choi and Heath were sponsored by the Texas Department of Transportation under Project 0-6877 entitled "Communications and Radar-Supported Transportation Operations and Planning (CAR-STOP)".

upgrading their network. The area coverage of Gbps connectivity of an MGB system makes it possible for cellular operators or broadband service providers to deploy large number of Wi-Fi access points to meet data traffic demands before 5G arrives. With a scalable Gbps infrastructure already in place by the time 5G arrives, cellular operators and broadband service providers can smoothly ramp up the deployment of 5G by adding onto or upgrading some of the existing small cells. The MGB system can even be used for other applications that may require high data rates. For example, it seems ideal to power the infrastructure of vehicle-to-infrastructure networks that will empower cars with gigabit connections and will enable connected and autonomous driving.

In summary, MGB provides a solution to the important backhaul problem in current and future cellular systems. In the duration of this paper, we explain the MGB system concept at a high level. Then we describe antenna and transceiver designs for the MGB system. Based on the proposed system, we provide simulation results to validate the claims made about the ability of MGB to support enough capacity. We find that the MGB system, in a reasonable system simulation setting is able to provide more than ten gigabits of backhaul capacity in a small cell setting.

## II. THE MILLIMETER-WAVE GBPS BROADBAND SYSTEM

Operators deploy cells with different sizes such as macro cells, micro cells, pico cells, and femto cells, which are generally connected to the core network via wired or wireless links. Intermediate nodes such as in-band relay stations have also been added, but the deployment is limited because the relay link competes with access links for the expensive licensed spectrum. Point-to-multi-point (PtMP) systems have also been used for backhaul, mostly using unlicensed spectrum in 5 GHz band, and the licensed spectrum in 28 GHz. To deliver area coverage, however, the hub station typically employs antennas with wide azimuth beamwidth, resulting in limited antenna gain and range.

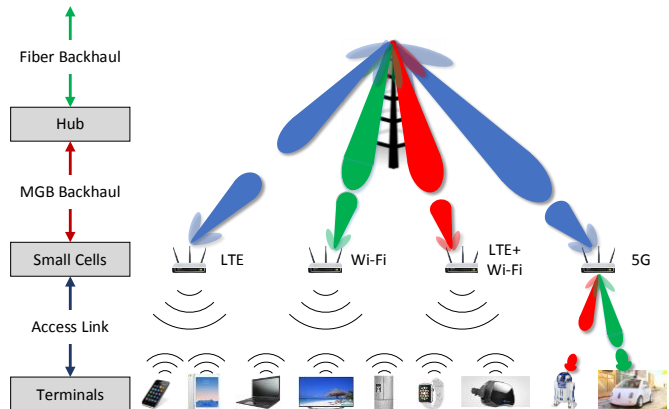

Fig. 1. A MGB system with LTE small cells, Wi-Fi access points, and 5G

Recent developments in phased antenna arrays make it possible to use antenna arrays with electronically steerable beams in commercial systems. For example, 16 phased array RF channels have been integrated into a single IC [4]. With phased array antennas, beams with high antenna gain can be formed and pointed to virtually any direction within a wide solid angle defined by the antenna pattern of the individual antenna elements. This not only can extend the range and coverage, but also reduce interference with other nearby transmissions.

The proposed MGB system is illustrated in Fig. 1. The system consists of a hub and a number of small cells. Communication between the hub and the small cells is established in millimeter wave spectrum ("MGB backhaul" as shown in the figure). Both the transmitter and the receiver use a hybrid of analog beamforming and digital MIMO processing to adapt to channel condition and balance analog power consumption and digital processing complexity. A typical MGB hub has 3 sectors. Each sector uses a planar phased antenna array and dynamically form beams to transmit to and receive from small cells. The small cells also use planar phased antenna arrays to point to the best directions to transmit to and receive from the MGB hub. On the access link, small cells can use either LTE, or Wi-Fi, or 5G, or the combination of these access technologies to communicate with a variety of devices including smart phones, tablets, laptops, TVs, home appliances, wearable devices, and future connected devices such as robots or self-driving vehicles.

With multiple small cells under its coverage, the MGB hub can use multi-user MIMO to communicate with multiple small cells simultaneously when needed. In addition, Time Division Multiple access (TDMA) and Orthogonal Frequency Division Multiple Access (OFDMA) are also supported in multiplexing traffic to and from small cells. For its primary purpose of boosting mobile or fixed broadband throughput, Time Division Duplex (TDD) is the preferred duplex scheme because its flexibility to adapt to the asymmetry of uplink and downlink data traffic, and the possibility to exploit channel reciprocity to adapt transmission schemes to maximize throughput.

The link budget for an exemplary MGB system at 39 GHz with 500 MHz system bandwidth and cell radius of 1 km is shown in Table 1.

Table 1. Link budget for an MGB system at 39 GHz

| 39 GHz MGB link budget | Downlink cell edge | Uplink cell edge | Downlink 4-Stream | Uplink 4-Stream |
|---|---|---|---|---|
| PA output (dBm) | 10 | 10 | 10 | 10 |
| # of PAs | 64 | 16 | 64 | 16 |
| Total power (dBm) | 28 | 22 | 28 | 22 |
| **EIRP (dBm)** | **55.14** | **43.10** | **55.14** | **43.10** |
| Distance (m) | 1000 | 1000 | 707 | 707 |
| **Total path loss (dB)** | **139.26** | **139.26** | **131.86** | **131.86** |
| Received power (dBm) | -84.12 | -96.16 | -76.71 | -88.75 |
| Bandwidth (MHz) | 500 | 500 | 500 | 500 |
| Noise Figure (dB) | 5.00 | 5.00 | 5.00 | 5.00 |
| # of MIMO streams | 1 | 1 | 4 | 4 |
| Receiver loss (dB) | 3.00 | 3.00 | 3.00 | 3.00 |

| Spectral efficiency | 5.34 | 3.44 | 15.45 | 15.45 |
| --- | --- | --- | --- | --- |
| **Throughput (Mbps)** | 2,668 | 1,720 | 7,725 | 7,725 |

The MGB hub uses a 256-element antenna array and 64 power amplifiers (PAs) with 10 dBm output power each. The small cell uses a 64-element antenna array and 16 PAs with 10 dBm output power each. Effective Isotropic Radiation Power (EIRP) of 55 dBm and 43 dBm can be achieved for downlink and uplink, respectively. The path loss is modeled by free space loss plus an additional loss of 15 dB per km to account for other factors such as rain, reflection, foliage, etc. More than 1 Gbps can be achieved in both the downlink and uplink at the cell edge (1 km from the hub). With 4-stream multi-user MIMO to 4 small cells with median path loss (707 meter from the hub), 7.7 Gbps system throughput can be achieved per sector in both the downlink and uplink.

Note that in the 39 GHz band 15 dB/km additional path loss corresponds to 60 mm/hr rainfall, which occurs less than 0.01% of the time (<1 hour per year) for most part of the world ([6]). The MGB transceivers at small cells should preferably be placed in locations where a good link (*e.g.*, line-of-sight or near-line-of-sight) can be established. In the worst case scenario when a small cell placed at the cell edge in a non-line-of-sight condition encounters heavy rain fall, the backhaul throughput of that small cell needs to be reduced. For example, reducing the throughput to 100 Mbps would provide another 20 dB margin beyond the 15 dB/km already included in the link budget calculation.

The OFDM numerology of a wide-area mmWave system needs to be carefully chosen, with considerations on coherence time, coherent bandwidth, and clock accuracy (frequency offset, phase noise, etc.), among others (See more detailed discussion in [1]). Fortunately, the relatively stable channel conditions in MGB systems mean large coherent bandwidth and ample time for synchronization, channel estimation, and beam tracking. This allows a wide range of OFDM numerology, providing flexibility to accommodate other implementation considerations. As a rule of thumb, we recommend the subcarrier spacing to be in the range of 100 kHz ~ 1 MHz, and the cyclic prefix to be in the range of 100 ns ~ 1 us. Alternatively, single-carrier waveform with similar cyclic prefix can be considered. But this makes it harder to more flexibly allocate resources as in OFDMA.

Adaptive modulation and coding are supported with modulation from QPSK to 64QAM and a variable code rate to allow rate adaptation based on channel conditions. We refrain from supporting 256QAM due to the stringent requirement on transceiver linearity (Error Vector Magnitude < -30 dB), which could significantly increase the cost of the transceiver design but only have limited usage in practice. We use low density parity check (LDPC) codes as the forward error correction (FEC) coding scheme to achieve Gbps decoding throughput with low power. LDPC codes have already been used in other mmWave standards like IEEE 802.11ad [7]. Practical design of LDPC decoders today can achieve multi-Gbps throughput with less than 100mW power consumption [8].

*Deployment of MGB systems*

By providing area coverage of Gbps connections, MGB systems significantly increase the site availability for small cells. This in turn lowers the site acquisition and site development cost, making dense deployment of small cells practical. In addition, the flexibility in Gbps backhaul access can potentially benefit the access network as small cells can be set up in locations that provide the best access link coverage without the limitation of wired backhaul.

The radius of an MGB system needs to be large enough to provide sufficient coverage, yet small enough to provide Gbps connectivity with great availability. We recommend the MGB cell radius to be in the range of 300 m – 3 km, achieving excellent economics in Gbps network deployment. A cell radius greater than 3 km leads to significant degradation of performance at the cell edge from the performance predicted by the link budget analysis in Table 1. A cell radius smaller than 300 m provides little footprint for Gbps backhaul for small cells, making it difficult to justify the cost of the system. Note this range of cell radius leads to deployment density comparable with the typical micro base station deployment density as in urban and suburban areas, allowing potential site sharing with existing mobile broadband infrastructure which can further lower the cost of the network. For example, for a MGB cell radius of 1 km, less than 500 hubs are needed to cover the whole New York City (a total area of 1214 $km^2$ with 8.4 million population).

The MGB hub should preferably be installed in towers and rooftops with similar requirement as base station sites (10 – 30 meter antenna height, fiber access, power, etc.) The MGB transceiver of small cells preferably should be installed at sufficient height, e.g., wall-mounted on the outside of a building or house. Although not required, it will be advantageous to choose a good location and general pointing direction for the MGB transceiver of a small cell so that a strong link with the hub can be established.

One of the main challenges in point-to-point microwave / millimeter-wave links is the installation of the bulky dish antennas. Due to its required high gain to close the link, the antennas are typically large and heavy, requiring heavy mounting equipment and skilled technicians onsite to complete the installation and calibration. Moreover, with large number of small cells deployed, maintenance cost also increases. With the MGB backhaul solution, both the hubs and the small cells have the ability to electronically steer the beams towards the most favorable direction, making it possible to automate a significant portion of the configuration and management of the system.

*Statistical multiplexing of small cell traffic*

The small footprint of small cells means fewer users per cell and increases the traffic variation among small cells and over time. For example, the traffic going through a small cell in a popular restaurant can be high during lunch or dinner while falling drastically in other hours. In addition, new data applications (e.g., Snapchat, Instagram, etc.) also tends to generate more bursty traffic. With 4G delivers hundreds of Mbps peak rate over the access link, it is expected 5G will provide Gbps access. The backhaul for each small cell therefore needs to provision for Gbps peak rate. However, the average

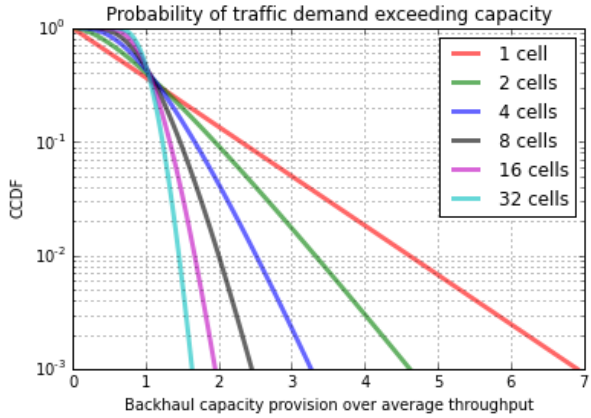

Fig. 2. Statistical multiplexing of small cell traffic reduce backhaul throughput fluctuation and the backhaul capacity headroom needed

throughput from each small cell is likely much smaller.

It is a tremendous challenge and huge waste to use wired, or point-to-point wireless solutions to provide backhaul for a large network of small cells. These solutions are not only very costly because operators need to deploy millions of small cells to providing meaningful upgrade to their network, but also highly inefficient because significant capacity headroom need to be provisioned to handle the Gbps peak rate, albeit it may only be utilized for a small fraction of time. The point-to-multi-point nature of the MGB system provides inherent statistical multiplexing of traffic from multiple small cells within its coverage, and thus providing the ability to handle Gbps peak rate from each small cell and the aggregated multi-Gbps throughput within the coverage of an MGB system in the most scalable and cost efficient manner.

Fig. 2 illustrates the backhaul capacity headroom needed to support small cells with bursty traffic. Assuming traffic from small cells is independent and follows the same exponential distribution, in order to meet the traffic demand 99% of the time without buffering, the backhaul capacity needs to be 4.6 times of the average throughput for a single cell, while only 1.46 times of the average throughput for 32-cell aggregation. Even when congested, the MGB system will have a lot more flexibility to schedule packets from different small cells, thus further mitigate the impact of backhaul delay. In other words, to provide the quality of service as described above to 32 small cells with 100 Mbps average throughput and 500 Mbps peak rate, network operator can either use 32 500-Mbps point-to-point backhaul links (coax cable, fiber, or wireless), or a single MGB sector. With multiple sectors an MGB hub can support even more small cells. This presents a significant scalability and cost advantage for MGB over coax cable, fiber, or point-to-point wireless backhaul solutions, especially in large scale deployment of small cells in dense urban areas.

## III. MGB Antenna Array And Transceiver

In this section we describe some potential antenna array designs and transceiver approaches in more detail.

Patch antenna arrays are one option for implementing antenna arrays with large number of elements in millimeter-wave frequencies with low cost. For frequencies in the range of 10 – 100 GHz, the dimension of the antenna elements are on the order of a few millimeters, which is easy to manufacture on printed circuit boards (PCB). This significantly brings down the cost of antenna arrays, compared with the conventional horn antennas or dish antennas. However, the loss of antenna feed network on PCB limits the dimension of the antenna arrays on PCB. For example, a carefully designed micro strip line on the best available PCB material can suffer up to 1 dB/inch of loss for frequency around 20 – 40 GHz. This effectively limit the size of the antenna array that we can make on PCB to a few hundreds elements for 20 – 40 GHz, with significant feed network loss expected for larger antenna arrays and even higher frequencies.

To achieve sufficient range, the antenna arrays at hub or small cell need to be sufficiently large. For example, in an exemplary system, the small cell antenna array has 64 antenna elements. With the build-in directivity of 6 dB for a patch antenna element, the total achievable directivity is 24 dB.

If each one of the 64 antenna elements is individually fed, the antenna array can freely scan the range within the antenna pattern of an individual element. That flexibility, however, comes with the high cost of 64 complete RF chains. To reduce the cost and complexity of RF transceiver, it is typically prudent to keep the number of RF chains at a minimum, while yet maintaining most of the capability for the MGB transceivers to dynamically form beams and adapt the MIMO processing schemes according to channel condition. This is done mainly through three techniques: antenna sub-arrays, analog beamforming, and digital MIMO processing.

An example of antenna sub-arrays is shown in Fig. 3(a). The 64 antenna elements are arranged in 8×8 fashion. Since most of the MGB hubs and small cells will be deployed less than 30 meters above the ground, it is OK to give up most of the beam steering capability in elevation. As such, the 64 antenna elements are grouped in to 16 4×1 sub-arrays. The 4 antenna elements within each sub-array are fed the same mmWave signals and form a fixed beam. As such, only 16 RF signals will be needed to feed this 64-element antenna array, resulting a 4x reduction of complexity of the RF circuits.

With analog beamforming, the complexity of the transceiver can be further reduced. Instead of converting all 16 channels of RF signals to baseband (which would require 16 I/Q de/modulator and 32 ADC/DACs), multiple channels of RF signals can be combined so that the number of Intermediate Frequency (IF) and Baseband (BB) channels is further reduced.

Fig. 3(b) shows an example of 4-channel mmWave front end with analog beamforming. The 4 mmWave channels are combined and frequency-converted to a single IF or Baseband channel. Each mmWave channel is equipped with two phase shifters (one for Tx, one for Rx). Analog beamforming is achieved by setting the phase of the phase shifters. In doing so, although we still have 4 mmWave channels, only a single IF & BB channel is needed to digitize the combined signal from them, resulting in significant savings of IF gain blocks, I/Q modulators/demodulators, ADC/DACs, and digital MIMO processing complexity.

As the number of RF chains gets larger, the amount of power needed from each RF channel gets smaller, making it possible to integrate multiple components of an RF chain, and possibly multiple RF chains, into a single integrated circuits. For example, all the circuit blocks shown in Fig. 3(b) can be integrated into a single mmWave front end IC. As shown in Fig. 3(b), only 4 of these mmWave FE ICs are needed to drive the 64-element antenna array.

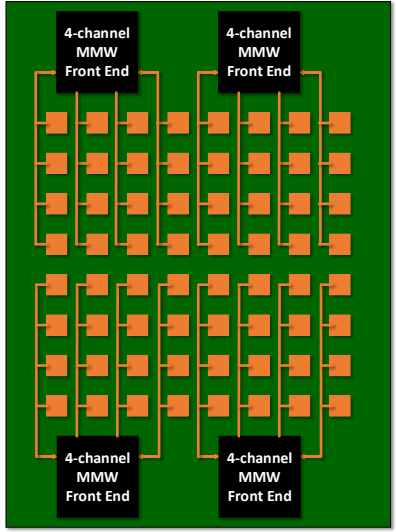 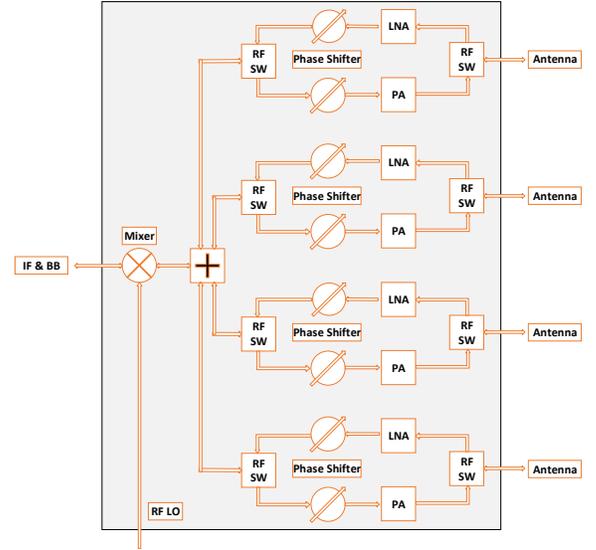

(a)          (b)

Fig. 3. An MGB Antenna Array and mmWave Front End IC

With 4x complexity reduction using 4×1 antenna subarrays, and 4x complexity reduction using 4-to-1 analog beamforming, the digital MIMO processor only needs to handle 4 MIMO streams to drive the 64-element antenna array, which is well within the capability of the advanced MIMO processors today. The joint analog beamforming and digital MIMO processing is sometimes called hybrid spatial processing or hybrid precoding and combing [7]. The analog beamformer is responsible for forming beams to adapt to long-term large-scale spatial channel characteristics such as angle of arrival (AoA) and angle of departure (AoD). The adaptation of analog beamformers in MGB system can be slow (on the order of 100 ms) due to the relatively static channel. However, even after analog beamforming towards the strongest spatial directions, near-field scattering (particularly around the small cells in non LOS conditions) can still occur, which increase the spatial resolution. The digital MIMO processor is responsible for adapt to the fast fading spatial channel given the analog beamforming. The digital MIMO processor need to adapt its equalizer per OFDM symbol or per transmission slot (around 10 – 100 us). Studies show that with proper antenna design and algorithm, hybrid spatial processing can achieve near optimal performance [9].

The thermal profile is also a significant factor in designing MGB transceivers. The amount of heat generated dictates the heat dissipation scheme and size of the heat dissipation device, which is a significant contributor to the size and weight of the hubs and small cells. It should be recognized that the solid state power amplifiers at millimeter-wave frequencies are not yet as efficient as their counterparts in lower frequencies. In order to maintain the thermal profile to be comparable to cellular or Wi-Fi transceivers, the transmission power needs to be reduced in comparison with conventional cellular base stations. Fortunately, the additional antenna gain at both the transmitter and the receiver can compensate for the higher path loss, and reduction of transmission power. This is shown in the link budget analysis in Table 1. In addition, as the power amplifiers are distributed on the PCB, the heat sources are distributed, mitigating the thermal challenge of MGB devices.

## IV. PERFORMANCE EVALUATION

We perform system level simulations to evaluate the proposed MGB solution. An MGB system with 19 hubs is simulated. The hubs are configured with 3 sectors per hub and arranged on a hexagonal grid with 1 km coverage radius (1.73 km inter-hub distance). Small cells are randomly dropped in the coverage area of the system for 10,000 simulation runs. The detailed simulation parameters are listed in Table 2.

Table 2. Simulation parameters

| Parameter | Assumption |
| --- | --- |
| Layout | Hexagonal |
| Hubs | 19 hubs, 3 sectors per hub |
| Hub radius (m) | 1000 |
| # of antennas at hub (per sector) | 256 (16×16 patch antenna array) |
| # of antennas at small cell | 64 (8×8 patch antenna array) |
| Total transmit power (dBm) | 28 |
| Antenna pattern | 3GPP model [10] |
| Min. distance b/w hub and small cell (m) | 100 |
| Carrier frequency $f_c$ (GHz) | 39 |
| System bandwidth (MHz) | 500 |
| Path loss model (dB) | $64.26+20\log_{10}(d)+0.015d$ (with distance $d$ in meter) |

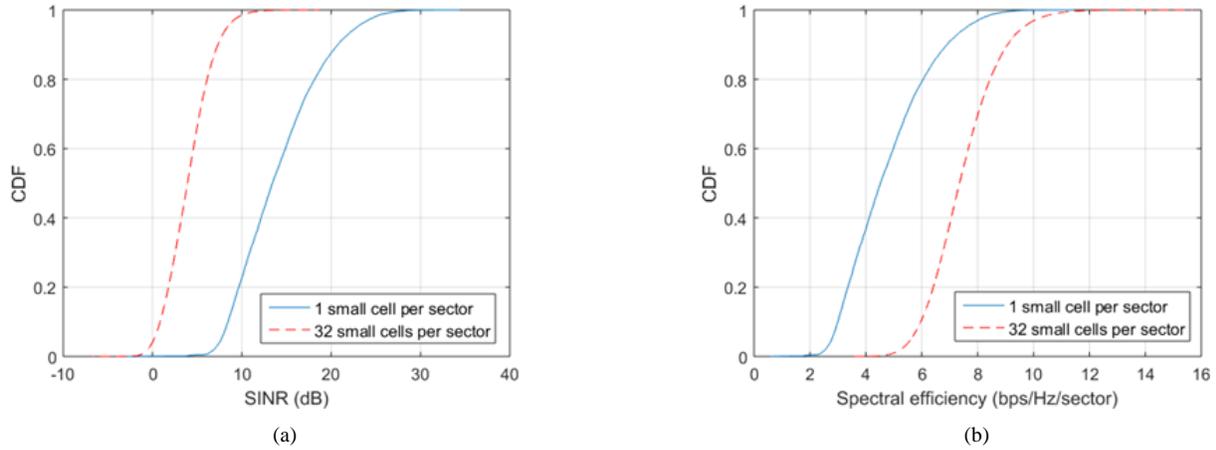

(a)                                        (b)

Fig. 4. SINR and Spectral Efficiency of the MGB system

| Noise figure (dB) | 5 |
|---|---|
| Receiver implementation loss (dB) | 3.00 |

For the first numerical study, we consider one small cell per sector. The cumulative distributed function (cdf) plots of signal-to-interference-noise ratio (SINR) and spectral efficiency are shown in Figs. 6(a) and 6(b), respectively. The average spectral efficiency for this case is 4.75 bps/Hz, or 2.38 Gbps throughput for each small cell. Among all the randomly dropped small cells, more than 99.6% achieve >2 bps/Hz spectral efficiency, or 1 Gbps throughput. These results can be viewed as the peak throughput per small cell when all resources of a sector are dedicated to a single small cell.

Next we consider the scenario with multiple small cells per sector. We assume each sector supports 32 small cells. To simplify small cell scheduling, in each time slot each sector transmits to 4 small cells that experience minimal spatial interference. The cdf plots of SINR and spectral efficiency of this study are also shown in Figs. 6(a) and 6(b), respectively. In this case, the average spectral efficiency becomes 7.46 bps/Hz per sector, which corresponds to 11.18 Gbps average throughput per hub. The harmonic mean throughput of 99% of the small cells (excluding 1% of the small cells with the lowest throughput) is 106.1 Mbps, indicating that MGB system can provide 99% guarantee of 100 Mbps average throughput with 32 small cells per sector (96 small cells per hub) via equal grade of service scheduling.

## V. CONCLUSION

In this paper, we propose a millimeter-wave Gbps broadband (MGB) system for both fixed and mobile broadband evolution towards 5G. Due to its distinctive ability to provide *wide-area Gbps coverage*, the MGB system is flexible, scalable, and cost effective as the last mile solution for Gbps fixed broadband and as the small cell backhaul solution for Gbps mobile broadband. In an exemplary MGB system with 500 MHz bandwidth at 39 GHz band, our simulation shows that a single MGB hub with 3 sectors can guarantee 1 Gbps peak rate and 100 Mbps average throughput to 96 small cells within 1 km radius with 99% probability.

REFERENCES

[1] Zhouyue Pi; Khan, F., "An introduction to millimeter-wave mobile broadband systems," Communications Magazine, IEEE , vol.49, no.6, pp.101,107, June 2011
[2] Marzetta, T.L., "Noncooperative Cellular Wireless with Unlimited Numbers of Base Station Antennas," Wireless Communications, IEEE Transactions on , vol.9, no.11, pp.3590,3600, November 2010
[3] Bhushan, N.; Junyi Li; Malladi, D.; Gilmore, R.; Brenner, D.; Damnjanovic, A.; Sukhavasi, R.; Patel, C.; Geirhofer, S., "Network densification: the dominant theme for wireless evolution into 5G," Communications Magazine, IEEE , vol.52, no.2, pp.82,89, February 2014
[4] Jeong-Geun Kim; Dong-Woo Kang; Byung-Wook Min; Rebeiz, G.M., "A single-chip 36-38 GHz 4-element transmit/receive phased-array with 5-bit amplitude and phase control," Microwave Symposium Digest, 2009. MTT '09. IEEE MTT-S International , vol., no., pp.561,564, 7-12 June 2009
[5] Sarkar, Anirban; Greene, Kevin; Floyd, Brian, "A power-efficient 4-element beamformer in 120-nm SiGe BiCMOS for 28-GHz cellular communications," Bipolar/BiCMOS Circuits and Technology Meeting (BCTM), 2014 IEEE , vol., no., pp.68,71, Sept. 28 2014-Oct. 1 2014
[6] Recommendation ITU-R P.837-1, "Characteristics of Precipitation for Propagation Modelling", Geneva, 2013
[7] T. Rappaport, R. W. Heath, Jr., R. Daniels, J. N. Murdock, "Millimeter Wave Wireless Communications," Pearson Education, Inc. 2014.
[8] Meng Li; Naessens, F.; Debacker, P.; Raghavan, P.; Desset, C.; Min Li; Dejonghe, A.; Van der Perre, L., "An area and energy efficient half-row-paralleled layer LDPC decoder for the 802.11AD standard," Signal Processing Systems (SiPS), 2013 IEEE Workshop on , vol., no., pp.112,117, 16-18 Oct. 2013
[9] El Ayach, O.; Rajagopal, S.; Abu-Surra, S.; Zhouyue Pi; Heath, R.W., "Spatially Sparse Precoding in Millimeter Wave MIMO Systems," Wireless Communications, IEEE Transactions on, vol.13, no.3, pp.1499-1513, March 2014.
[10] 3GPP TR 25.996 v12.0.0, "Spatial channel model for Multiple Input Multiple Output (MIMO) simulations," Technical Report 3GPP, Sep. 2014.